\pdfoutput=1

\documentclass[12pt,a4paper]{article}

\usepackage{ifthen} 
\newboolean{pdflatex}
\setboolean{pdflatex}{true} 

\newboolean{articletitles}
\setboolean{articletitles}{true} 

\newboolean{uprightparticles}
\setboolean{uprightparticles}{false} 

\newboolean{inbibliography}
\setboolean{inbibliography}{false} 

\usepackage[top=1in, bottom=1.25in, left=1in, right=1in]{geometry}

\usepackage{microtype}
\usepackage{lineno}  
\usepackage{xspace} 
\usepackage{caption} 

\usepackage{graphicx}  
\usepackage{color}
\usepackage{colortbl}
\graphicspath{{./figs/}} 

\usepackage{amsmath} 
\usepackage{amssymb}
\usepackage{amsfonts}
\usepackage{upgreek} 

\newcommand*\patchAmsMathEnvironmentForLineno[1]{%
\expandafter\let\csname old#1\expandafter\endcsname\csname #1\endcsname
\expandafter\let\csname oldend#1\expandafter\endcsname\csname
end#1\endcsname
 \renewenvironment{#1}%
   {\linenomath\csname old#1\endcsname}%
   {\csname oldend#1\endcsname\endlinenomath}%
}
\newcommand*\patchBothAmsMathEnvironmentsForLineno[1]{%
  \patchAmsMathEnvironmentForLineno{#1}%
  \patchAmsMathEnvironmentForLineno{#1*}%
}
\AtBeginDocument{%
\patchBothAmsMathEnvironmentsForLineno{equation}%
\patchBothAmsMathEnvironmentsForLineno{align}%
\patchBothAmsMathEnvironmentsForLineno{flalign}%
\patchBothAmsMathEnvironmentsForLineno{alignat}%
\patchBothAmsMathEnvironmentsForLineno{gather}%
\patchBothAmsMathEnvironmentsForLineno{multline}%
\patchBothAmsMathEnvironmentsForLineno{eqnarray}%
}

\usepackage{hyperref}    
\usepackage[all]{hypcap} 

\usepackage{xspace} 
\usepackage{upgreek}

\def\lhcb {\mbox{LHCb}\xspace}

\def\lhc    {\mbox{LHC}\xspace}

\def\MagUp {\mbox{\em Mag\kern -0.05em Up}\xspace}

\ifthenelse{\boolean{uprightparticles}}%
{

    \def\Pmu         {\ensuremath{\upmu}\xspace}

    \def\Ppi         {\ensuremath{\uppi}\xspace}

    \def\Ptau        {\ensuremath{\uptau}\xspace}                 
                     
    \def\Pphi        {\ensuremath{\upphi}\xspace}

    \def\Ppsi        {\ensuremath{\uppsi}\xspace}

    \def\PDelta      {\ensuremath{\Delta}\xspace}                 
    \def\PXi      {\ensuremath{\Xi}\xspace}                 
    \def\PLambda      {\ensuremath{\Lambda}\xspace}                 
    \def\PSigma      {\ensuremath{\Sigma}\xspace}                 
    \def\POmega      {\ensuremath{\Omega}\xspace}                 
    \def\PUpsilon      {\ensuremath{\Upsilon}\xspace}                 
    
    \def\PB      {\ensuremath{\mathrm{B}}\xspace}                 
                     
    \def\PD      {\ensuremath{\mathrm{D}}\xspace}

    \def\PJ      {\ensuremath{\mathrm{J}}\xspace}                 
    \def\PK      {\ensuremath{\mathrm{K}}\xspace}

    \def\PP      {\ensuremath{\mathrm{P}}\xspace}

    \def\PS      {\ensuremath{\mathrm{S}}\xspace}

    \def\Pb      {\ensuremath{\mathrm{b}}\xspace}                 
    \def\Pc      {\ensuremath{\mathrm{c}}\xspace}                 
    \def\Pd      {\ensuremath{\mathrm{d}}\xspace}

    \def\Pi      {\ensuremath{\mathrm{i}}\xspace}

    \def\Pp      {\ensuremath{\mathrm{p}}\xspace}

    \def\Ps      {\ensuremath{\mathrm{s}}\xspace}                 
                     
    \def\Pu      {\ensuremath{\mathrm{u}}\xspace}

}
{

    \def\Pmu         {\ensuremath{\mu}\xspace}

    \def\Ppi         {\ensuremath{\pi}\xspace}

    \def\Ptau        {\ensuremath{\tau}\xspace}                 
                     
    \def\Pphi        {\ensuremath{\phi}\xspace}

    \def\Ppsi        {\ensuremath{\psi}\xspace}                 
                     
    \mathchardef\PDelta="7101
    \mathchardef\PXi="7104
    \mathchardef\PLambda="7103
    \mathchardef\PSigma="7106
    \mathchardef\POmega="710A
    \mathchardef\PUpsilon="7107
                     
    \def\PB      {\ensuremath{B}\xspace}                 
                     
    \def\PD      {\ensuremath{D}\xspace}

    \def\PJ      {\ensuremath{J}\xspace}                 
    \def\PK      {\ensuremath{K}\xspace}

    \def\PP      {\ensuremath{P}\xspace}

    \def\PS      {\ensuremath{S}\xspace}

    \def\Pb      {\ensuremath{b}\xspace}                 
    \def\Pc      {\ensuremath{c}\xspace}                 
    \def\Pd      {\ensuremath{d}\xspace}

    \def\Pi      {\ensuremath{i}\xspace}

    \def\Pp      {\ensuremath{p}\xspace}

    \def\Ps      {\ensuremath{s}\xspace}                 
                     
    \def\Pu      {\ensuremath{u}\xspace}

}

\makeatletter
\ifcase \@ptsize \relax
\newcommand{\miniscule}{\@setfontsize\miniscule{4}{5}}
\or
\newcommand{\miniscule}{\@setfontsize\miniscule{5}{6}}
\or
\newcommand{\miniscule}{\@setfontsize\miniscule{5}{6}}
\fi
\makeatother

\DeclareRobustCommand{\optbar}[1]{\shortstack{{\miniscule (\rule[.5ex]{1.25em}{.18mm})}
        \\ [-.7ex] $#1$}}

\def\muon       {{\ensuremath{\Pmu}}\xspace}
\def\mup        {{\ensuremath{\Pmu^+}}\xspace}
\def\mun        {{\ensuremath{\Pmu^-}}\xspace} 
\def\mumu       {{\ensuremath{\Pmu^+\Pmu^-}}\xspace}

\def\tauon      {{\ensuremath{\Ptau}}\xspace}

\def\uquark    {{\ensuremath{\Pu}}\xspace}

\def\dquark    {{\ensuremath{\Pd}}\xspace}

\def\squark    {{\ensuremath{\Ps}}\xspace}

\def\cquark    {{\ensuremath{\Pc}}\xspace}

\def\bquark    {{\ensuremath{\Pb}}\xspace}
\def\bquarkbar {{\ensuremath{\overline \bquark}}\xspace}
\def\bbbar     {{\ensuremath{\bquark\bquarkbar}}\xspace}

\def\pion   {{\ensuremath{\Ppi}}\xspace}

\def\pip    {{\ensuremath{\pion^+}}\xspace}

\def\kaon    {{\ensuremath{\PK}}\xspace}
\def\Kbar    {{\kern 0.2em\overline{\kern -0.2em \PK}{}}\xspace}

\def\KorKbar    {\kern 0.18em\optbar{\kern -0.18em K}{}\xspace}

\def\Kp      {{\ensuremath{\kaon^+}}\xspace}
\def\Km      {{\ensuremath{\kaon^-}}\xspace}

\def\Kstarz  {{\ensuremath{\kaon^{*0}}}\xspace}

\def\Kstar   {{\ensuremath{\kaon^*}}\xspace}

\newcommand{\phiz}{\ensuremath{\Pphi}\xspace}

\def\Dbar    {{\kern 0.2em\overline{\kern -0.2em \PD}{}}\xspace}
\def\D       {{\ensuremath{\PD}}\xspace}

\def\DorDbar    {\kern 0.18em\optbar{\kern -0.18em D}{}\xspace}
\def\Dz      {{\ensuremath{\D^0}}\xspace}

\def\Dstarp  {{\ensuremath{\D^{*+}}}\xspace}

\def\B       {{\ensuremath{\PB}}\xspace}
\def\Bbar    {{\ensuremath{\kern 0.18em\overline{\kern -0.18em \PB}{}}}\xspace}

\def\BorBbar    {\kern 0.18em\optbar{\kern -0.18em B}{}\xspace}

\def\Bu      {{\ensuremath{\B^+}}\xspace}

\def\Bd      {{\ensuremath{\B^0}}\xspace}
\def\Bs      {{\ensuremath{\B^0_\squark}}\xspace}

\def\jpsi     {{\ensuremath{{\PJ\mskip -3mu/\mskip -2mu\Ppsi\mskip 2mu}}}\xspace}
\def\psitwos  {{\ensuremath{\Ppsi{(2S)}}}\xspace}

\def\Y#1S{\ensuremath{\PUpsilon{(#1S)}}\xspace}

\def\proton      {{\ensuremath{\Pp}}\xspace}

\def\Lbar        {{\ensuremath{\kern 0.1em\overline{\kern -0.1em\PLambda}}}\xspace}
\def\LorLbar    {\kern 0.18em\optbar{\kern -0.18em \PLambda}{}\xspace}

\def\BF         {{\ensuremath{\mathcal{B}}}}

\def\BR         {\BF}
\newcommand{\decay}[2]{\ensuremath{#1\!\to #2}\xspace}         

\def\to                 {\ensuremath{\rightarrow}\xspace}

\def\order   {{\ensuremath{\mathcal{O}}}\xspace}

\def\Bq      {{\ensuremath{\B^0_{(\squark)}}}\xspace}

\def\BsToJPsiPhi  {\decay{\Bs}{\jpsi\phiz}}
\def\BdToJPsiKst  {\decay{\Bd}{\jpsi\Kstarz}}

\def\BToSmmPmm {\decay{\Bq}{\PS(\to\mumu)\PP(\to\mumu)}}
\def\BsToSmmPmm {\decay{\Bs}{\PS(\to\mumu)\PP(\to\mumu)}}
\def\BdToSmmPmm {\decay{\Bd}{\PS(\to\mumu)\PP(\to\mumu)}}

\def\BqToMuMuMuMu {\decay{\Bq}{\mumu\mumu}}
\def\BsToMuMuMuMu {\decay{\Bs}{\mumu\mumu}}
\def\BdToMuMuMuMu {\decay{\Bd}{\mumu\mumu}}
\def\Bmmmm {\BqToMuMuMuMu}
\def\Bqmmmm {\Bmmmm}
\def\Bsmmmm {\BsToMuMuMuMu}
\def\Bdmmmm {\BdToMuMuMuMu}
\def\BsToMuMuMuMu {\decay{\Bs}{\mumu\mumu}}

\def\BuToJpsiK    {\decay{\Bu}{\jpsi(\to\mumu)\Kp}}
\def\BsToJpsiPhimmmm  {\decay{\Bs}{\jpsi(\to\mumu)\phiz(\to\mumu)}}
\def\BsToJpsiPhiKK  {\decay{\Bs}{\jpsi(\to\mumu)\phiz(\to\Kp\Km)}}

\def\BqToSPmmmm       {\decay{\Bq}{S(\to\mumu)P(\to\mumu)}}

\def\BdToKstmm    {\decay{\Bd}{\Kstarz\mup\mun}}

\def\BdToJPsiKst  {\decay{\Bd}{\jpsi\Kstarz}}

\def\AT#1     {\ensuremath{A_{\mathrm{T}}^{#1}}\xspace}           

\def\C#1      {\ensuremath{\mathcal{C}_{#1}}\xspace}                       
\def\Cp#1     {\ensuremath{\mathcal{C}_{#1}^{'}}\xspace}                    
\def\Ceff#1   {\ensuremath{\mathcal{C}_{#1}^{\mathrm{(eff)}}}\xspace}        
\def\Cpeff#1  {\ensuremath{\mathcal{C}_{#1}^{'\mathrm{(eff)}}}\xspace}       
\def\Ope#1    {\ensuremath{\mathcal{O}_{#1}}\xspace}                       
\def\Opep#1   {\ensuremath{\mathcal{O}_{#1}^{'}}\xspace}                    

\newcommand{\tev}{\ensuremath{\mathrm{\,Te\kern -0.1em V}}\xspace}
\newcommand{\gev}{\ensuremath{\mathrm{\,Ge\kern -0.1em V}}\xspace}
\newcommand{\mev}{\ensuremath{\mathrm{\,Me\kern -0.1em V}}\xspace}
\newcommand{\kev}{\ensuremath{\mathrm{\,ke\kern -0.1em V}}\xspace}
\newcommand{\ev}{\ensuremath{\mathrm{\,e\kern -0.1em V}}\xspace}
\newcommand{\gevc}{\ensuremath{{\mathrm{\,Ge\kern -0.1em V\!/}c}}\xspace}
\newcommand{\mevc}{\ensuremath{{\mathrm{\,Me\kern -0.1em V\!/}c}}\xspace}
\newcommand{\gevcc}{\ensuremath{{\mathrm{\,Ge\kern -0.1em V\!/}c^2}}\xspace}
\newcommand{\gevgevcccc}{\ensuremath{{\mathrm{\,Ge\kern -0.1em V^2\!/}c^4}}\xspace}
\newcommand{\gevgevcc}{\ensuremath{{\mathrm{\,Ge\kern -0.1em V^2\!/}c^2}}\xspace}
\newcommand{\mevcc}{\ensuremath{{\mathrm{\,Me\kern -0.1em V\!/}c^2}}\xspace}

\def\mum  {\ensuremath{{\,\upmu\mathrm{m}}}\xspace}

\def\invfb   {\ensuremath{\mbox{\,fb}^{-1}}\xspace}

\newcommand{\stat}{\ensuremath{\mathrm{\,(stat)}}\xspace}
\newcommand{\syst}{\ensuremath{\mathrm{\,(syst)}}\xspace}

\def\order{{\ensuremath{\mathcal{O}}}\xspace}
\newcommand{\chisq}{\ensuremath{\chi^2}\xspace}

\newcommand{\chisqip}{\ensuremath{\chi^2_{\text{IP}}}\xspace}

\def\gsim{{~\raise.15em\hbox{$>$}\kern-.85em
        \lower.35em\hbox{$\sim$}~}\xspace}
\def\lsim{{~\raise.15em\hbox{$<$}\kern-.85em
        \lower.35em\hbox{$\sim$}~}\xspace}

\def\sqs   {\ensuremath{\protect\sqrt{s}}\xspace}

\def\ptot       {\mbox{$p$}\xspace}
\def\pt         {\mbox{$p_{\mathrm{ T}}$}\xspace}

\def\dllmupi    {\ensuremath{\mathrm{DLL}_{\muon\pi}}\xspace}

\def\evtgen     {\mbox{\textsc{EvtGen}}\xspace}

\def\geant      {\mbox{\textsc{Geant4}}\xspace}

\def\photos     {\mbox{\textsc{Photos}}\xspace}

\def\pythia     {\mbox{\textsc{Pythia}}\xspace}

\def\tell1  {TELL1\xspace}
\def\ukl1   {UKL1\xspace}

\usepackage{cite} 
\usepackage{mciteplus}

\usepackage{subfigure}

\usepackage{longtable} 

\begin{document}

\renewcommand{\thefootnote}{\fnsymbol{footnote}}
\setcounter{footnote}{1}


\begin{titlepage}
\pagenumbering{roman}

\vspace*{-1.5cm}
\centerline{\large EUROPEAN ORGANIZATION FOR NUCLEAR RESEARCH (CERN)}
\vspace*{1.5cm}
\noindent
\begin{tabular*}{\linewidth}{lc@{\extracolsep{\fill}}r@{\extracolsep{0pt}}}
\ifthenelse{\boolean{pdflatex}}
{\vspace*{-2.7cm}\mbox{\!\!\!\includegraphics[width=.14\textwidth]{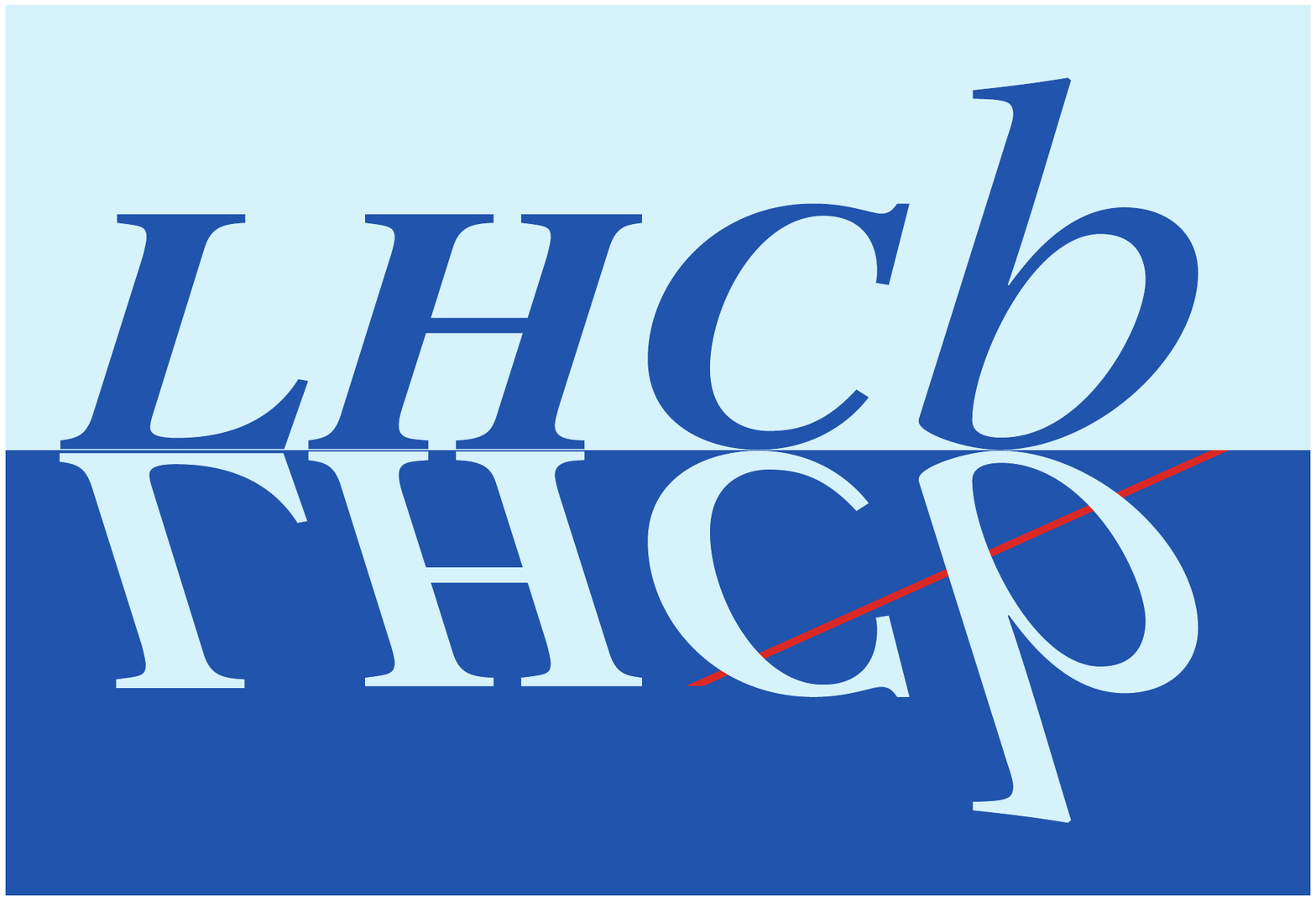}} & &}%
{\vspace*{-1.2cm}\mbox{\!\!\!\includegraphics[width=.12\textwidth]{lhcb-logo.eps}} & &}%
\\
 & & CERN-EP-2016-276 \\  
 & & LHCb-PAPER-2016-043 \\  
 & & March 1, 2017 \\ 
 & & \\
\end{tabular*}

\vspace*{4.0cm}

{\normalfont\bfseries\boldmath\huge
\begin{center}
  Search for decays of neutral beauty mesons into four muons
\end{center}
}

\vspace*{2.0cm}

\begin{center}
The LHCb collaboration\footnote{Authors are listed at the end of this paper.}
\end{center}

\vspace{\fill}

\begin{abstract}
  \noindent
  A search for the non-resonant decays \Bsmmmm and \Bdmmmm is presented.
  The measurement is performed using the full Run 1 data set collected in proton-proton collisions by the \lhcb experiment at the \lhc.
  The data correspond to integrated luminosities of 1 and 2\invfb collected at centre-of-mass energies of 7 and 8\tev, respectively.
  No signal is observed and upper limits on the branching fractions of the non-resonant decays at 95\% confidence level are determined to be
  \begin{align*}
     \BR(\Bsmmmm) & < 2.5 \times 10^{-9},\\ 
     \BR(\Bdmmmm) & < 6.9 \times 10^{-10}.
    \end{align*}

\end{abstract}

\vspace*{2.0cm}

\begin{center}
  Published in JHEP
\end{center}

\vspace{\fill}

{\footnotesize 
\centerline{\copyright~CERN on behalf of the \lhcb collaboration, licence \href{http://creativecommons.org/licenses/by/4.0/}{CC-BY-4.0}.}}
\vspace*{2mm}

\end{titlepage}


\newpage
\setcounter{page}{2}
\mbox{~}
%
%
%
%

\cleardoublepage


\renewcommand{\thefootnote}{\arabic{footnote}}
\setcounter{footnote}{0}



\pagestyle{plain} 
\setcounter{page}{1}
\pagenumbering{arabic}


%

\section{Introduction}
\label{sec:Introduction}

\begin{figure}[tbb]
    \centering
    \subfigure[\label{subfig:nonResoFMD}]{\includegraphics[width=0.32\textwidth]{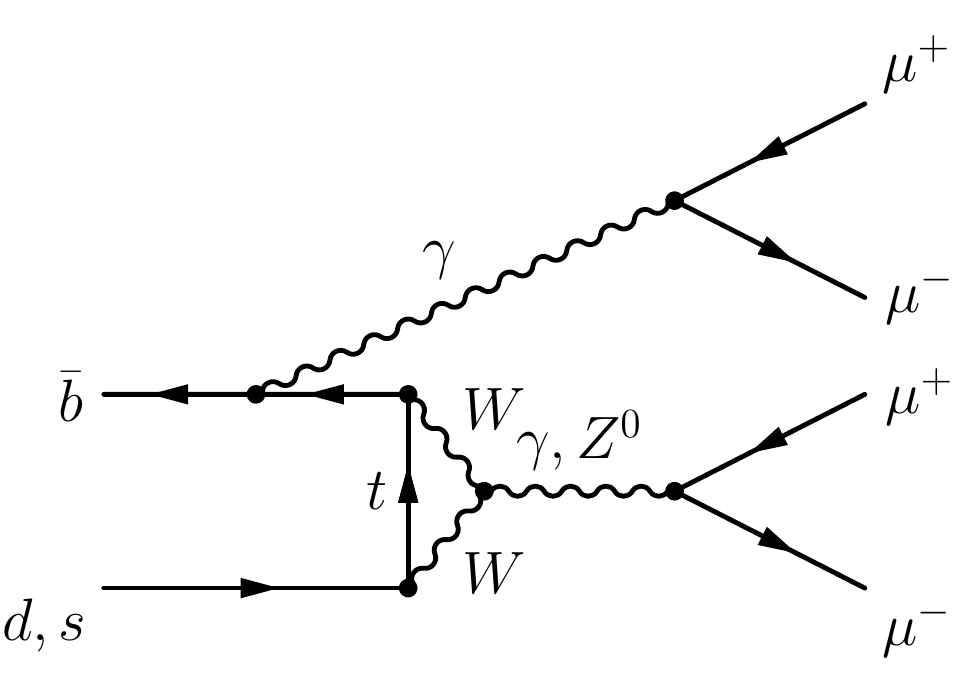}}
    \subfigure[\label{subfig:NPFMD}]{\includegraphics[width=0.32\textwidth]{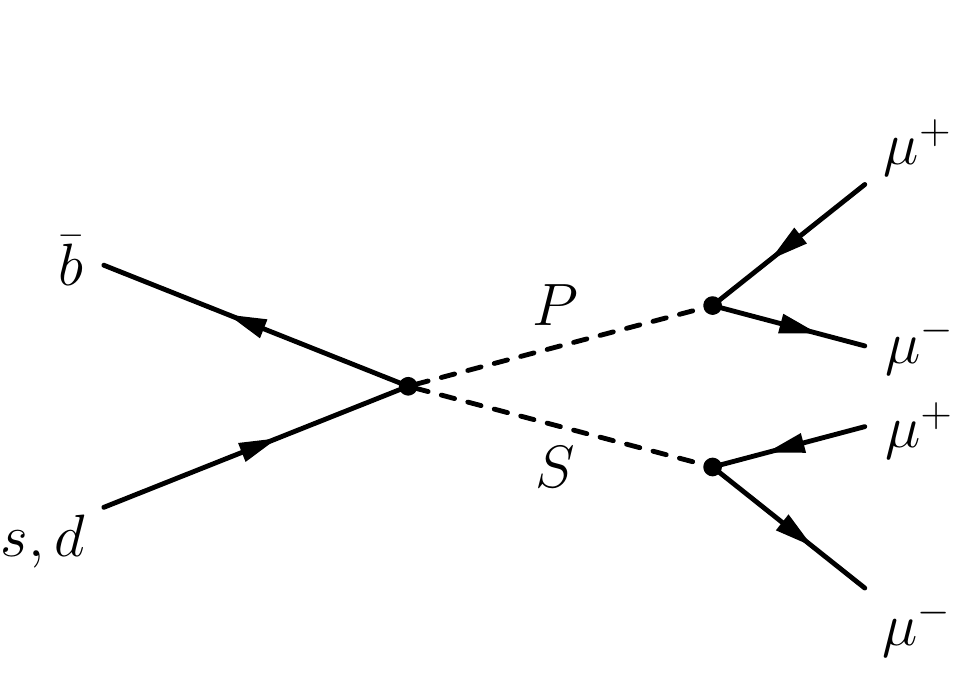}}
    \subfigure[\label{subfig:ResoFMD}]{\includegraphics[width=0.32\textwidth]{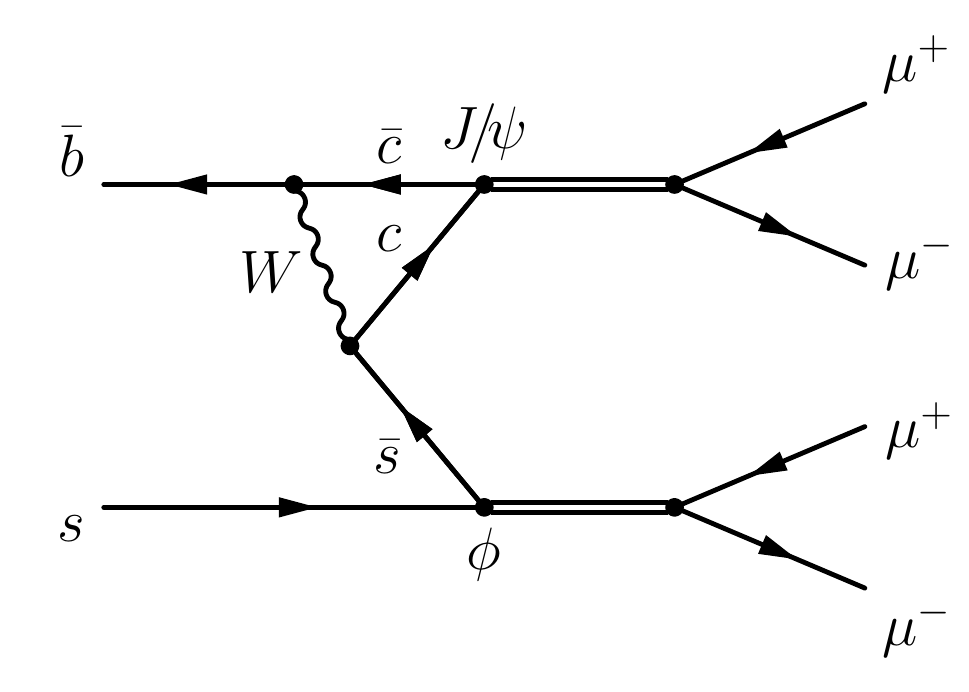}}\hspace{0.01\textwidth}
    \caption{Feynman diagrams for~\subref{subfig:nonResoFMD} the non-resonant \mbox{\Bqmmmm} decay,~\subref{subfig:NPFMD} a supersymmetric \mbox{\BToSmmPmm} decay and~\subref{subfig:ResoFMD} the resonant \protect\linebreak\BsToJpsiPhimmmm decay (see text).~\label{fig:feynmans}}
\end{figure}

The rare decays \Bqmmmm proceed through \decay{\bquark}{\dquark (\squark)} flavour-changing neutral-current processes, which are strongly suppressed in the Standard Model (SM).\footnote{The inclusion of charge-conjugate processes is implied throughout.}
In the main non-resonant SM amplitude, one muon pair is produced via amplitudes described by electroweak loop diagrams and the other is created by a virtual photon as shown in Fig.~\ref{subfig:nonResoFMD}.
The branching fraction of the non-resonant \Bsmmmm decay is expected to be  $3.5 \times 10^{-11}$\cite{Dincer:2003zq}.

Theories extending the SM can significantly enhance the \Bmmmm decay rate by contributions of new particles.
For example, in minimal supersymmetric models (MSSM), the decay can proceed via new scalar \PS and pseudoscalar \PP sgoldstino particles, which both decay into a dimuon final state as shown in Fig.~\ref{subfig:NPFMD}.
There are two types of couplings between sgoldstinos and SM fermions.
Type-I couplings describe interactions between a sgoldstino and two fermions, where the coupling strength is proportional to the fermion mass.
Type-II couplings describe a four-particle vertex, where a scalar and a pseudoscalar sgoldstino interact with two fermions.
Branching fractions up to $\BR(\decay{\Bs}{\PS\PP}) \approx 10^{-4}$ and $\BR(\decay{\Bd}{\PS\PP}) \approx 10^{-7}$ are possible~\cite{Demidov:2011rd}.
Sgoldstinos can decay into a pair of photons or a pair of charged leptons~\cite{Demidov:2006pt}.
In this analysis the muonic decay is considered, as the coupling to electrons is smaller and the large \tauon-lepton mass limits the available phase space.ṣ
The branching fractions of the sgoldstino decays strongly depend on the model parameters such as the sgoldstino mass and the supersymmetry breaking scale.
In the search for \decay{\PSigma^{+}}{\proton\mumu} decays the HyperCP collaboration found an excess of events, which is consistent with the decay \decay{\PSigma^{+}}{\PP\proton} with \decay{\PP}{\mumu} and a pseudoscalar mass of $m(\PP) = 214.3\pm0.5\mev$~\cite{Park:2005eka}.

So far only limits on the SM and MSSM branching fractions at 95\% confidence level have been measured by \lhcb based on the data recorded in 2011~\cite{LHCb-PAPER-2012-049} to be
\begin{align*}
\BR(\Bsmmmm) &< 1.6 \times 10^{-8},\\
\BR(\Bdmmmm) &< 6.6 \times 10^{-9},\\
\BR(\BsToSmmPmm) &< 1.6 \times 10^{-8},\\
\BR(\BdToSmmPmm) &< 6.3 \times 10^{-9}.
\end{align*}
These limits are based on assumed sgoldstino masses of \mbox{$m(\PS) = 2.5\gevcc$}, which is approximately the central value of the allowed mass range, and \mbox{$m(\PP) = 214.3\mevcc$}.

The dominant SM decays of neutral \B mesons into four muons proceed through resonances like \phiz, \jpsi and \psitwos.
The most frequent of these decays is \BsToJPsiPhi, where both the \jpsi and the \phiz mesons decay into a pair of muons, as shown in Fig.~\ref{subfig:ResoFMD}.
In the following, this decay is referred to as the resonant decay mode and treated as a background.
From the product of the measured branching fractions of the underlying decays \BR(\BsToJPsiPhi), \BR(\decay{\jpsi}{\mumu}), and \BR(\decay{\phiz}{\mumu})~\cite{PDG2016} its branching fraction is calculated to be $(1.83 \pm 0.18) \times 10^{-8}$.

In this paper a search for the non-resonant SM process, and for the MSSM-induced \Bmmmm decays is presented, using \proton\proton collision data recorded by the \lhcb detector during \lhc Run 1.
Potentially contributing sgoldstinos are assumed to be short lived, such that they do not form a displaced vertex.
The analysed data correspond to integrated luminosities of 1 and 2\invfb collected at centre-of-mass energies of 7 and 8\tev, respectively.
The branching fraction is measured relative to the decay \mbox{\BuToJpsiK}, which gives a clean signal with a precisely measured branching fraction~\cite{PDG2016}. 
This yields a significant improvement compared to the previous measurement, where the use of the \BdToJPsiKst decay as normalisation mode resulted in a large systematic uncertainty originating from the S-wave fraction and the less precisely measured branching fraction.
The advantage of normalising to a well-known \B meson decay is that dominant systematic uncertainties originating mainly from the \bbbar cross-section cancel in the ratio.

\section{Detector and Simulation}
\label{sec:detector_and_simulation}

The \lhcb detector~\cite{Alves:2008zz,LHCb-DP-2014-002} is a single-arm forward spectrometer covering the \mbox{pseudorapidity} range $2<\eta <5$, designed for the study of particles containing \bquark or \cquark quarks.
The detector includes a high-precision tracking system consisting of a silicon-strip vertex detector surrounding the $pp$ interaction region
, a large-area silicon-strip detector located upstream of a dipole magnet with a bending power of about $4{\mathrm{\,Tm}}$, and three stations of silicon-strip detectors and straw drift tubes 
placed downstream of the magnet.
The tracking system provides a measurement of momentum, \ptot, of charged particles with a relative uncertainty that varies from 0.5\% at low momentum to 1.0\% at 200\gevc.
The minimum distance of a track to a primary vertex (PV), the impact parameter, is measured with a resolution of $(15+29/\pt)\mum$, where \pt is the component of the momentum transverse to the beam, in\,\gevc.
Different types of charged hadrons are distinguished using information from two ring-imaging Cherenkov (RICH) detectors.
Photons, electrons and hadrons are identified by a calorimeter system consisting of scintillating-pad and preshower detectors, an electromagnetic calorimeter and a hadronic calorimeter.
Muons are identified by a system composed of alternating layers of iron and multiwire proportional chambers. 

In the simulation, $pp$ collisions are generated using~\pythia~\cite{Sjostrand:2006za,Sjostrand:2007gs} with a specific \lhcb configuration~\cite{LHCb-PROC-2010-056}.
Decays of hadronic particles are described by \evtgen~\cite{Lange:2001uf}, in which final-state radiation is generated using \photos~\cite{Golonka:2005pn}.
The interaction of the generated particles with the detector, and its response, are implemented using the \geant toolkit~\cite{Allison:2006ve, Agostinelli:2002hh} as described in Ref.~\cite{LHCb-PROC-2011-006}.

\section{Event selection}
\label{sec:event_selection}

The online event selection is performed by a trigger
, which consists of a hardware stage, based on information from the calorimeter and muon systems, followed by a software stage, which applies a full event reconstruction.
In this analysis candidate events are first required to pass the hardware trigger, which for 7\tev (8\tev) data selects events with at least one muon with a transverse momentum of $\pt>1.48\gevc$ 
($\pt>1.76\gevc$) or at least one pair of muons with the product of the transverse momenta larger than $(1.296)^2$ \gevgevcc\linebreak ($(1.6)^2$ \gevgevcc).
In the subsequent software trigger, at least one of the final-state particles is required to have $\pt>1\gevc$ and an impact parameter larger than $100\mum$ with respect to all PVs in the event.

In the offline selection, the \Bq decay vertex is constructed from four good quality muon candidates that form a common vertex and have a total charge of zero.
The vertex is required to be significantly displaced from any PV.
Among the four final-state muons, there are four possible dimuon combinations with zero charge.
In all four combinations, the mass windows corresponding to the \phiz $(950 - 1090\mevcc)$, \jpsi $(3000 - 3200\mevcc)$ and \psitwos $(3600 - 3800\mevcc)$ resonances are vetoed.
This efficiently suppresses any background from any of the three mentioned resonances to a negligible level.
Contributions of other charmonium states are found to be negligible.

The MatrixNet (MN)~\cite{MatrixNet}, a multivariate classifier based on a Boosted Decision Tree~\cite{Breiman,AdaBoost}, is applied in order to remove combinatorial background, where a candidate \Bq vertex is constructed from four muons that do not originate from a single \B meson decay.
The input variables are the following properties of the \Bq candidate: the decay time, the vertex quality, the momentum and transverse momentum, the cosine of the direction angle (DIRA), and the smallest impact parameter chisquare (\chisqip) with respect to any PV, where \chisqip is defined as the difference between the vertex-fit \chisq of a PV reconstructed with and without the \Bq candidate. 
The DIRA is defined as the angle between the momentum of the reconstructed \Bq candidate and the vector from the PV with the smallest \chisqip to the \Bq decay vertex.
As training samples, simulated \Bsmmmm and \Bdmmmm events, generated with a uniform probability across the decay phase space, are used as a signal proxy.
Before training, the signal simulation is weighted to correct for known discrepancies between data and simulation as described later.
The background sample is taken from the far and the near sidebands in data as defined in Table~\ref{tab:masswindows}.
In order to verify that the classification of each event is unbiased, 10-fold cross-validation~\cite{Blum:1999:BHB:307400.307439} is employed.

Background arising from misidentifying one or more particles is suppressed by applying particle identification (PID) requirements.
Information from the RICH system, the calorimeters and the muon system is used to calculate the difference in log-likelihood between the hypothesis of a final-state particle being a pion or a muon, \dllmupi.

Events in the signal region are not examined until the analysis is finalised.
Events outside the signal region are split into the far sidebands, used to calculate the expected background yield, and the near sidebands, used to optimise the cuts on the MN response and the minimum \dllmupi values of the four muon candidates in the final state. 
The optimization of the cuts is performed on a two-dimensional grid maximising the figure of merit~\cite{Punzi:2003bu}
\begin{table}[tbp]
    \caption{Definitions of intervals in the \Bd and \Bs reconstructed invariant mass distributions.~\label{tab:masswindows}}
    \centering
    \begin{tabular}{lc}
    	                        &    Mass interval (\mevcc)     \\ \hline
    	Near sidebands          &  $[5020,5220]$ and $[5426,5626]$   \\
    	Far sidebands           &  $[4360,5020]$ and $[5626,6360]$   \\
    	Signal region           & $[m(\Bd)-60,m(\Bs) + 60]$ \\
    	\Bs search region       &       $[m(\Bs)-40,m(\Bs) + 40]$   \\
    	\Bd search region       &      $[m(\Bd)-40,m(\Bd) + 40]$
    \end{tabular}
\end{table}
\begin{align*}
    \mathrm{FoM}&=\frac{\varepsilon_\mathrm{signal}}{\sigma/2+\sqrt{N_\mathrm{bkg}^\mathrm{expected} \times \varepsilon_\mathrm{bkg}}}.
\end{align*}
The intended significance in terms of standard deviations ($\sigma$) is set to three.
Very similar selection criteria are found when using five.
The expected background yield before applying the MN and PID selection, $N_\mathrm{bkg}^\mathrm{expected}$, is determined from a fit to the events in the near sidebands using an exponential function.
For each grid point the background efficiency, $\varepsilon_\mathrm{bkg}$, is measured using events from the near sidebands.
The signal efficiency, $\varepsilon_\mathrm{signal}$, is measured for each grid point using simulated \Bmmmm decays.
Lacking a model for non-resonant \Bqmmmm simulation, the selection of the preceding measurement was developed on \BsToJpsiPhimmmm data. Now that a suitable simulation model is available, significant improvements in terms of signal efficiency and background rejection are made by employing a multivariate classifier and being able to measure the selection efficiency from simulation.

\section{Selection efficiencies and systematic uncertainties}
\label{sec:eff_and_sys}

The optimal working point corresponds to signal efficiencies of $({0.580 \pm 0.003})\%$ and $({0.568 \pm 0.003})\%$ for the \mbox{\Bsmmmm} and \mbox{\Bdmmmm} decay modes, respectively.
Sources of peaking background such as \mbox{\BdToKstmm}, in which the kaon and the pion originating from the \Kstar decay are misidentified as muons, are reduced to a negligible level by the optimised selection.
The efficiencies for the MSSM processes are measured using simulated samples of the \mbox{\BqToSPmmmm} decays, where the \Bq meson decays into a pseudoscalar sgoldstino with a mass of 214.3\mevcc~\cite{Park:2005eka} and a scalar sgoldstino with a mass of 2.5\gevcc.
Both the \PP and \PS particles are simulated with a decay width of $\Gamma = 0.1\mevcc$, which corresponds to a prompt decay.
The measured efficiencies are the same for the \Bs and the \Bd decays and amount to $({0.648 \pm 0.003})\%$. 
The difference between the SM and the MSSM efficiencies originates from the fact that in the case of the decay proceeding via \PP and \PS sgoldstinos, the decay products are more likely to be within the acceptance of the \lhcb detector.
In order to test the dependence of the measured \BToSmmPmm branching fractions on the mass of the scalar sgoldstino, the selection efficiency is measured in bins of dimuon invariant mass while requiring the corresponding other dimuon mass to be between 200 and 950\mevcc.
An efficiency variation of $\order(20\%)$ is observed.  

The selection applied to the normalisation mode \BuToJpsiK differs from that applied to the signal modes in the PID criteria and that no multivariate analysis technique is applied.
The total efficiency is $({1.495 \pm 0.006})\%$.
The uncertainties on the efficiencies are driven by the limited number of simulated events and are treated as systematic uncertainties of 0.4 -- 0.5\%.

The total efficiency is calculated as the product of the efficiencies of the different stages of the selection.
As an alternative to the trigger efficiency calculated on simulation, the value is calculated on \BuToJpsiK data~\cite{LHCb-PUB-2014-039} and a systematic uncertainty of 3\% is assigned corresponding to the relative difference. 
The efficiency of the MN classifier to select the more frequent decay \BsToJpsiPhiKK is compared between data and simulation.
The relative difference of 0.3\% is assigned as a systematic uncertainty.
Another source of systematic uncertainty arises from the track finding efficiency.
Again, values obtained from data~\cite{LHCb-DP-2013-002} and simulation are compared and the deviation is treated as a correction factor for the efficiency, while the uncertainty on the deviation, 1.7\%, is assigned as a systematic uncertainty.

In general the agreement in the observables used in the selection between data and simulation is very good, although there are some distributions that are known to deviate.
Therefore, the gradient boosting reweighting technique~\cite{gbreweighter} is used to calculate weights that correct for differences between data and simulation in \BsToJpsiPhiKK.
The weighting is performed in the track multiplicity, the \B transverse momentum, the \chisq of the decay vertex fit and the \chisqip.
The first two are chosen because they are correlated with the PID variables and the latter two dominate the feature ranking obtained from the MN training.
These weights are applied to the \Bmmmm and \BuToJpsiK simulation samples, and are used to calculate the MN and the PID efficiencies.
In order to account for inaccuracies of this method resulting from the kinematic and topological differences between the decay modes, systematic uncertainties of 3.6\% are assigned based on the difference of the MN efficiency on \Bqmmmm and \BsToJpsiPhiKK.
For the \BuToJpsiK decay mode, the efficiencies are measured with and without weights and the observed difference of 2.3\% is assigned as systematic uncertainty.

In order to determine accurate efficiencies of the applied PID requirements, calibration samples of muons from \decay{\jpsi}{\mumu} and \decay{\phiz}{\mumu} decays and of kaons from \decay{\Dstarp}{\Dz(\to\Km\pip)\pip} decays are used.
The relative frequency for kaons and muons to pass the PID criteria is calculated in bins of track multiplicity, particle momentum and pseudorapidity.
Different binning schemes are tested and the observed differences in the efficiencies of 1\% for \BuToJpsiK and 0.5\% for \Bqmmmm are assigned as systematic uncertainties.
Additionally, 3\% of the simulated \Bmmmm decays contain muons with low transverse momentum outside the kinematic region covered by the calibration data.
This fraction is assigned as a systematic uncertainty.
Candidates that have a reconstructed invariant mass within $\pm 40\mevcc$ around the known \Bq mass, which corresponds to $\pm2\sigma$ of the mass resolution, are treated as signal candidates.
The accuracy of the efficiency of this cut is evaluated on \BsToJpsiPhiKK data.
A systematic uncertainty of 0.5\% corresponding to the relative difference of the efficiency measured on data and simulation is assigned.
Systematic uncertainties of 0.9\% and 0.5\% in the case of \Bqmmmm and \BuToJpsiK originate from the imperfections of the efficiency of the event reconstruction due to soft photon radiation and 0.6\% from mismatching of track segments between different tracking stations in the detector, which is measured using simulated events.
All relevant sources of systematic uncertainty along with the total values are summarised in Table~\ref{tab:systSummary}.
The most significant improvements with regard to the preceding measurement are the larger available data sample, and the choice of the \BuToJpsiK decay as normalisation mode, which has the advantage of a precisely measured branching fraction and the absence of an additional systematic uncertainty originating from the S-wave correction.

\section{Normalisation}
\label{sec:event_yields}
\begin{figure}[htpb]
    \centering
    \includegraphics[width=0.7\textwidth]{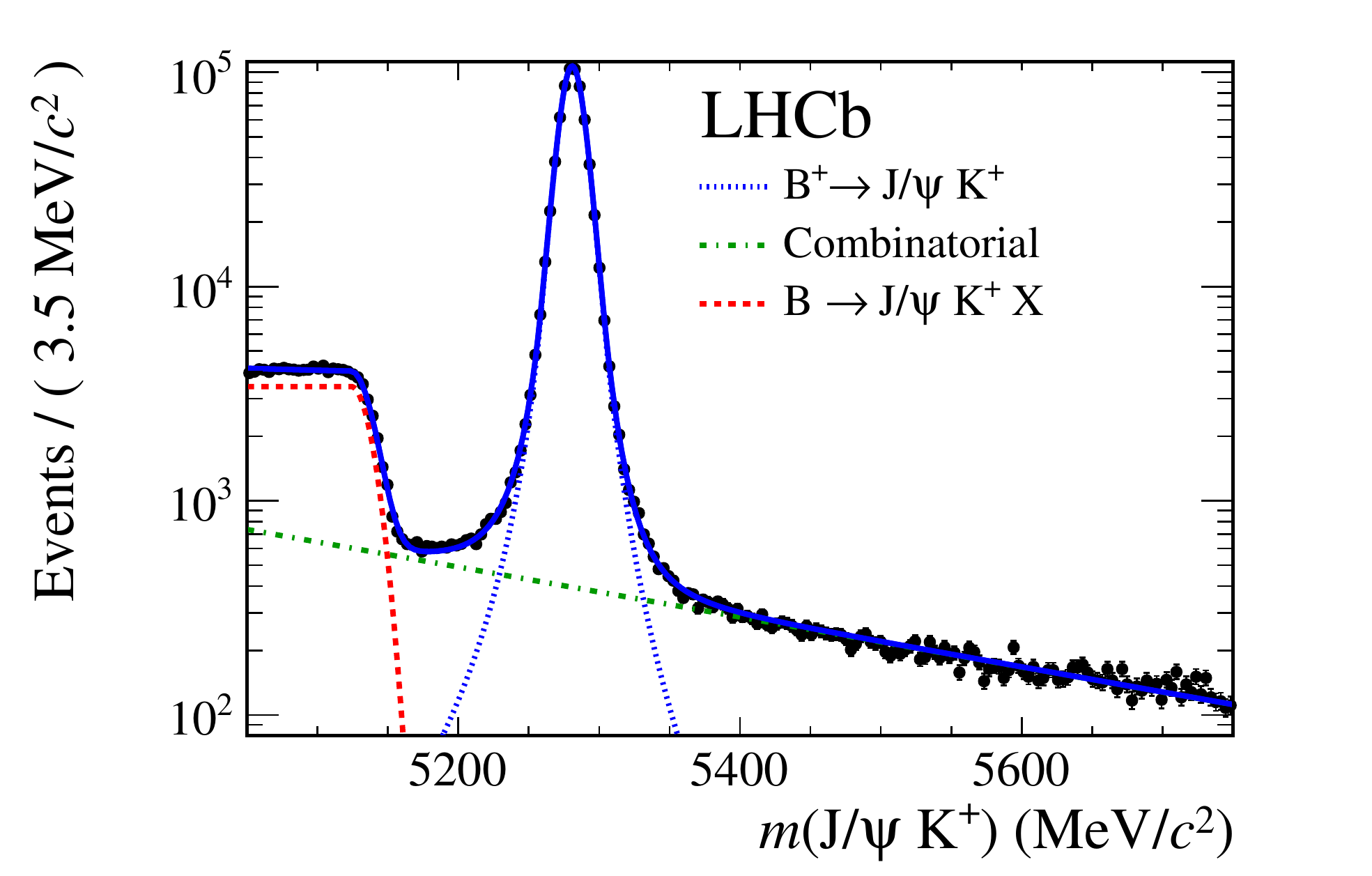}
    \caption{Fit to the \BuToJpsiK invariant mass distribution. The signal contribution is modelled by a Hypatia2~\cite{ipatia} function (blue dotted line), the combinatorial background by an exponential function (green dash-dotted line). Partially reconstructed decays, such as \BdToJPsiKst where one pion is not reconstructed, are modelled by a Gaussian function with an exponential tail towards the lower mass side (red dashed line). Data are shown by black dots.~\label{fig:jpsikFit}}
\end{figure}

The \BuToJpsiK signal yield is determined by performing an unbinned extended maximum likelihood fit to the \Kp\mumu invariant mass distribution.
In this fit the \jpsi mass is constrained~\cite{Hulsbergen:2005pu} to the world average~\cite{PDG2016}.
The normalisation yield is found to be $N(\BuToJpsiK) = 687890 \pm 920$.
The \jpsi\Kp mass spectrum along with the fit result is shown in Fig.~\ref{fig:jpsikFit}.
A systematic uncertainty of 0.3\% is assigned to the determined \BuToJpsiK yield by using an alternative fit model and performing a binned extended maximum likelihood fit.

The \Bmmmm branching fraction is calculated as
\begin{align*}
    \BR(\BqToMuMuMuMu) &= N(\BqToMuMuMuMu) \times \eta_{\dquark,\squark},
\end{align*} 
with
\begin{align*}
    \eta_{\dquark,\squark} &=  \frac{\varepsilon(\BuToJpsiK)\times\BR(\BuToJpsiK)}{\varepsilon(\BqToMuMuMuMu)\times N(\BuToJpsiK)} \times \frac{f_u}{f_{\dquark,\squark}},
\end{align*}
where $N(\BuToJpsiK)$ and $N(\BqToMuMuMuMu)$ are the observed yields of the normalisation and the signal channel, respectively.
The ratio between the production rates of \Bs and \Bd was measured by \lhcb to be $f_\squark/f_\dquark = 0.259 \pm 0.015$~\cite{fsfd}.
The measurement was performed using \proton\proton collision data at \sqs = 7\tev, but found to be stable between \sqs = 7\tev and 8\tev by a previous \lhcb measurement~\cite{LHCB-PAPER-2013-046}.
The ratio between the \Bu and \Bd production rates is assumed to be unity.
As a consequence $f_\squark/f_\uquark$ is equal to $f_\squark/f_\dquark$.

The single event sensitivities, $\eta_{\dquark,\squark}$, amount
\begin{align*}
\eta_s^\mathrm{SM} &= (8.65 \pm 0.80)\times 10^{-10},\\
\eta_d^\mathrm{SM} &=  (2.29 \pm 0.16)\times 10^{-10},\\
\eta_s^\mathrm{MSSM} &= (7.75 \pm 0.72)\times 10^{-10},\\
\eta_d^\mathrm{MSSM} &=  (2.01 \pm 0.14)\times 10^{-10},
\end{align*}
for the \Bs and the \Bd decay modes in the SM and in the MSSM scenario.
Here, the uncertainties are the combined values of the statistical uncertainty on the \linebreak \BuToJpsiK  yield and the systematic uncertainty.
In the case of $\eta_s$ the systematic uncertainty is dominated by the ratio of $f_u / f_s$ and in the case of $\eta_d$ by the weighting procedure applied to correct for the difference between data and simulation.

The individual sources of systematic uncertainties given in Table~\ref{tab:systSummary} are assumed to be uncorrelated and are combined quadratically.
The total systematic uncertainty is $9.2\%$ for the \Bs decay and $7.2\%$ for the \Bd decay.
These values are small compared to the statistical uncertainty on the expected number of background events in the \Bd and \Bs search regions.
The whole analysis strategy is cross-checked by measuring the \BsToJpsiPhimmmm branching fraction.
The obtained value has a precision of 20\% and is compatible with the product of the branching fractions of the underlying decays taken from Ref.~\cite{PDG2016}.

\begin{table}[h]
    \centering
    \caption{Summary of systematic uncertainties affecting the single event sensitivities along with the total systematic uncertainty calculated by adding up the individual components in quadrature. The dominating uncertainty arising from $f_\uquark / f_\squark$ only contributes to $\eta_s$. The uncertainty of the stated selection efficiencies arising from the limited number of simulated events is 0.5\% for \Bdmmmm and 0.4\% for all other considered decay modes.~\label{tab:systSummary}}
        \begin{tabular}{lc}
        	Source                                     & Value [\%] \\ \hline
        	Selection efficiency                       & $0.4-0.5$  \\
        	Trigger efficiency                         &    3.0     \\
        	MN efficiency                              &    0.3     \\
        	Track finding efficiency                   &    1.7     \\
        	Weighting \Bqmmmm                          &    3.6     \\
        	Weighting \BuToJpsiK                       &    2.3     \\
        	PID binning \BuToJpsiK                     &    1.0     \\
        	PID binning \Bmmmm                         &    0.5     \\
        	Kinematic coverage of PID calibration data &    3.0     \\
        	$\pm$40\mevcc search region efficiency     &    0.5     \\
        	Soft photon radiation \Bmmmm               &    0.9     \\
        	Soft photon radiation \BuToJpsiK           &    0.5     \\
        	Track segments mismatching                 &    0.6     \\
        	Normalisation fit                          &    0.3     \\
        	$f_\uquark / f_\squark$                    &    5.8     \\
        	\BR(\decay{\Bu}{\jpsi\Kp})                 &    3.0     \\
        	\BR(\decay{\jpsi}{\mumu})                  &    0.1     \\ \hline
        	Combined $\eta_s$ SM                       &    9.2     \\
        	Combined $\eta_d$ SM                       &    7.2     \\
        	Combined $\eta_s$ MSSM                     &    9.2     \\
        	Combined $\eta_d$ MSSM                     &    7.2
        \end{tabular}
\end{table}

The number of expected background events is determined by fitting an exponential function to the far sidebands of $m(\mumu\mumu)$.
Extrapolating and integrating the fitted function in $\pm 40\mevcc$ wide windows around the \Bq meson masses yields the number of expected background events,
\begin{align*}
    N_\mathrm{bkg}^\mathrm{expected}(\Bd) &= 0.55^{+0.24}_{-0.19}\stat\pm0.20\syst \quad\mathrm{and}\\
    N_\mathrm{bkg}^\mathrm{expected}(\Bs) &= 0.47^{+0.23}_{-0.18}\stat\pm0.18\syst.
\end{align*}
The statistical uncertainty is the combination of the Poissonian uncertainty originating from the limited size of the data sample and the uncertainty on the fit parameters.
As an alternative fit model a second-order polynomial is used and the difference between these background expectations is assigned as a systematic uncertainty.

\section{Results}
\label{sec:results}

\begin{figure}[btp]
	\centering
	\includegraphics[width=0.9\linewidth]{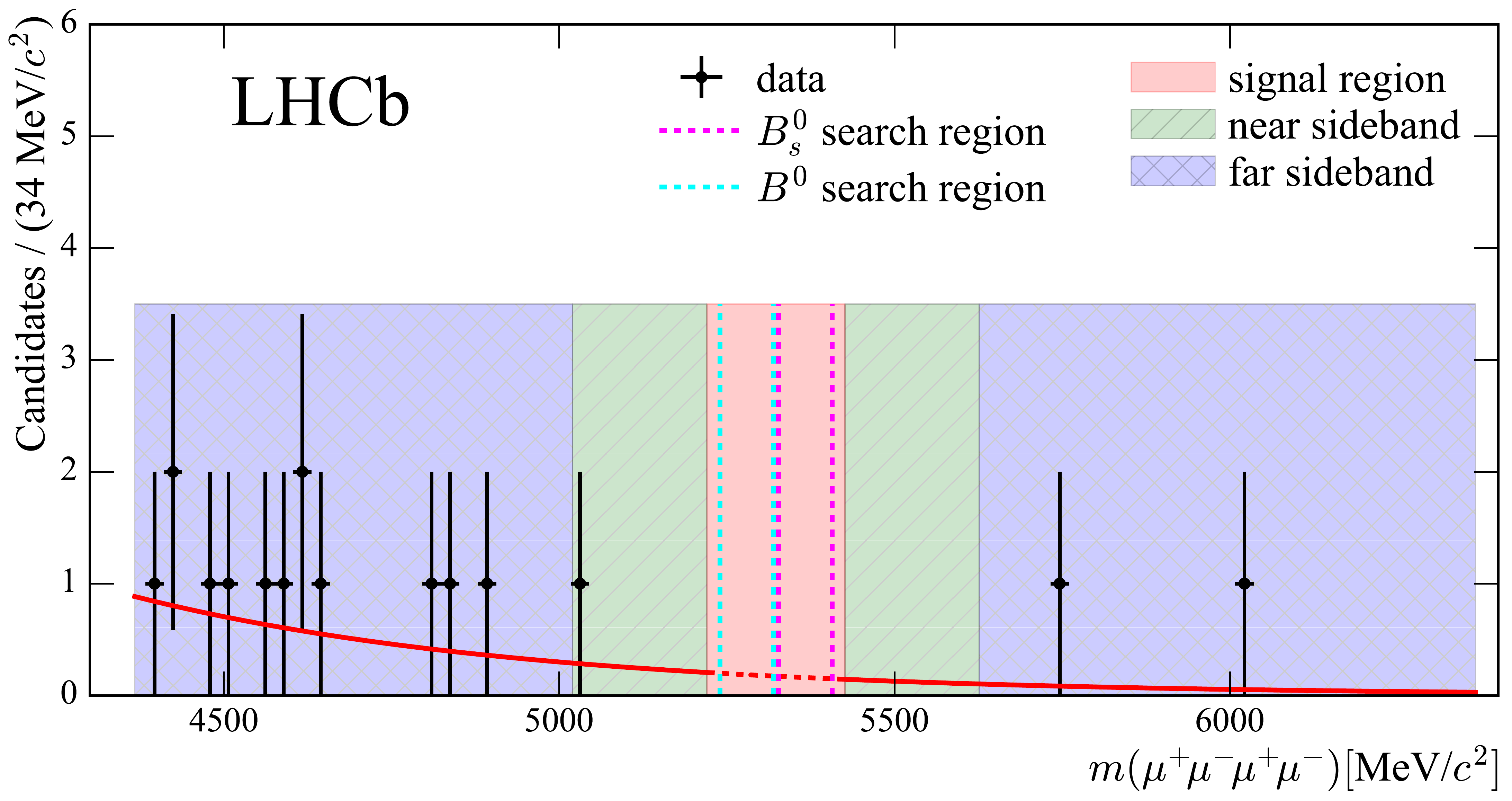}
	\caption{Mass distribution of selected \Bmmmm events observed in 3\invfb of data in all considered \B mass regions. Background (red line) is modelled by an exponential function. Signal subregions for \Bd and \Bs searches are also shown. The error bars on the individual 
		points with $n$ entries are $\pm\sqrt{n}$.~\label{fig:fit_to_B24mu}}
\end{figure}

The final distribution of the reconstructed mass of the four muon system is shown in Fig.~\ref{fig:fit_to_B24mu}.
No candidates are observed in either the \Bd or the \Bs search region, which is consistent with the expected background yield. 

\begin{figure}[h] 
    \begin{center}
        \includegraphics[width=0.6\linewidth]{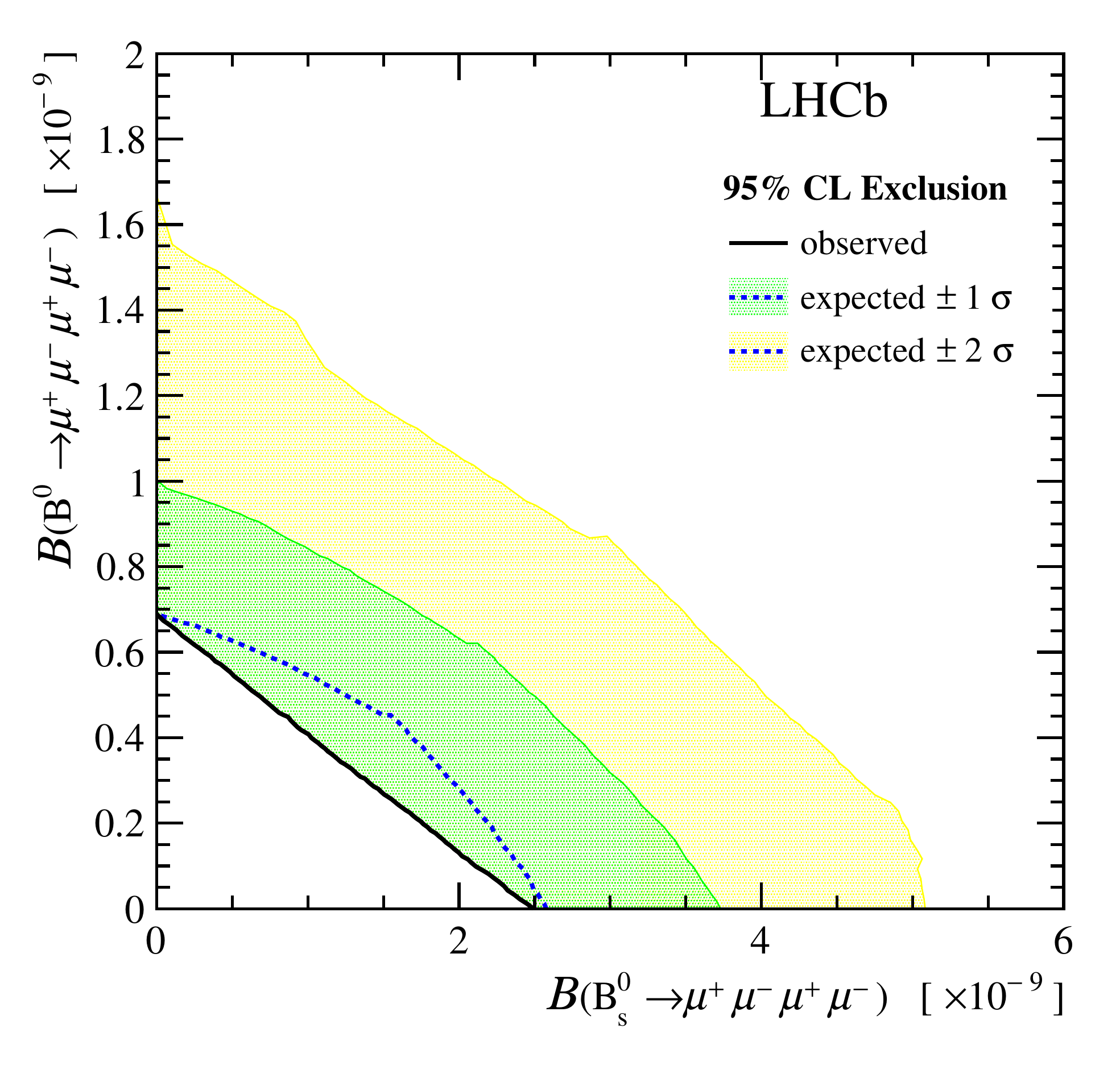} 
    \end{center}
    \caption{Expected and observed 95\% CL exclusion in \mbox{\BR(\Bdmmmm)} vs.\protect\linebreak \BR(\Bsmmmm) parameters plane.}
    \label{fig:cls_br_bd_bs}
\end{figure}

\begin{figure}[h] 
 \begin{center}
     \includegraphics[width=0.6\linewidth]{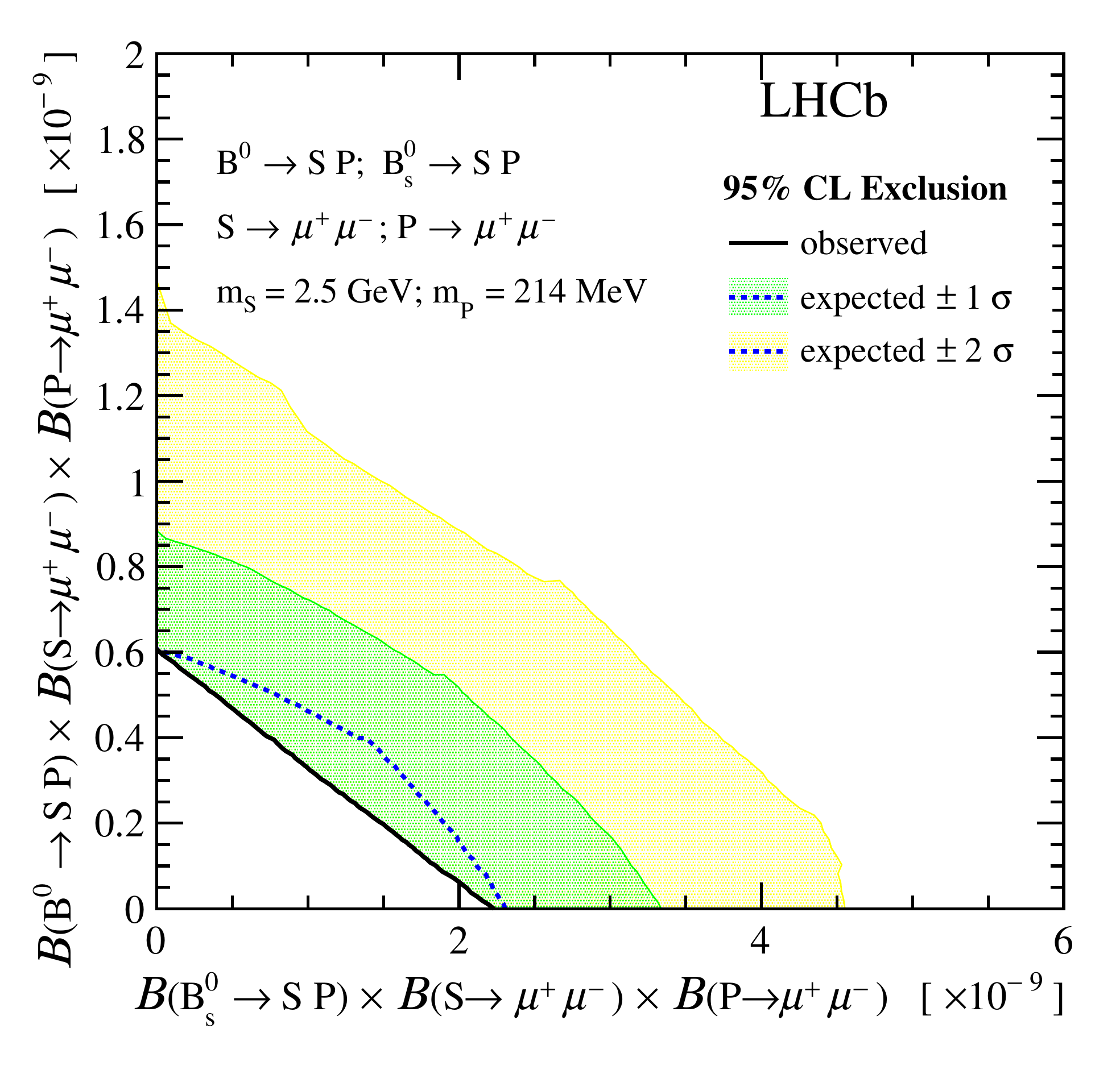}
 \end{center}
\caption{Expected and observed 95\% CL exclusion in \BR(\BdToSmmPmm) vs.\protect\linebreak \BR(\BsToSmmPmm)
  parameters plane with scalar and pseudoscalar S and P as described in Sec. \ref{sec:event_selection}.}
\label{fig:cls_br_bd_bsnp}
\end{figure}

The Hybrid CLs procedure~\cite{Cousins:1991qz,Read:2002hq,Junk:1999kv}, with log-normal priors 
to account for uncertainties of both background and efficiency estimations, is used to convert the 
observations into upper limits on the corresponding branching fractions. 
The exclusion at 95\% confidence level assuming the SM single event sensitivities is shown in Fig.~\ref{fig:cls_br_bd_bs}. 
The result for the corresponding MSSM values is presented in Fig.~\ref{fig:cls_br_bd_bsnp}.
The limits on the branching fractions of the \Bs and \Bd decays are anti-correlated.
Replacing the log-normal priors by gamma distributions yields the same results.

Assuming negligible cross-feed between the \Bs and the \Bd search regions, the observed upper limits on the branching fractions at 95\% confidence level are found to be
\begin{align*}
    \BR(\Bsmmmm) & < 2.5 \times 10^{-9}, \\ 
    \BR(\Bdmmmm) & < 6.9 \times 10^{-10},\\
    \BR(\BsToSmmPmm) &  < 2.2 \times 10^{-9},\\
    \BR(\BdToSmmPmm) & < 6.0 \times 10^{-10}.
\end{align*}

\section{Conclusion}
\label{sec:summary}
In summary, a search for non-resonant \Bmmmm decays has been presented.
In addition, the sensitivity to a specific MSSM scenario has been probed.
The applied selection focuses on finding four muon tracks that form a common vertex.
For the SM scenario and the MSSM decay through short-lived scalar and pseudoscalar new particles,
the limits set by the previous measurement performed by \lhcb on a subset of the present data, are improved by a factor of $6.4$ $(7.3)$ for the SM (MSSM) mode in the case of the \Bs decay and by a factor of $9.5$ $(10.5)$ in the case of the \Bd decay.



\section*{Acknowledgements}

 
\noindent We express our gratitude to our colleagues in the CERN
accelerator departments for the excellent performance of the LHC. We
thank the technical and administrative staff at the LHCb
institutes. We acknowledge support from CERN and from the national
agencies: CAPES, CNPq, FAPERJ and FINEP (Brazil); NSFC (China);
CNRS/IN2P3 (France); BMBF, DFG and MPG (Germany); INFN (Italy); 
FOM and NWO (The Netherlands); MNiSW and NCN (Poland); MEN/IFA (Romania); 
MinES and FASO (Russia); MinECo (Spain); SNSF and SER (Switzerland); 
NASU (Ukraine); STFC (United Kingdom); NSF (USA).
We acknowledge the computing resources that are provided by CERN, IN2P3 (France), KIT and DESY (Germany), INFN (Italy), SURF (The Netherlands), PIC (Spain), GridPP (United Kingdom), RRCKI and Yandex LLC (Russia), CSCS (Switzerland), IFIN-HH (Romania), CBPF (Brazil), PL-GRID (Poland) and OSC (USA). We are indebted to the communities behind the multiple open 
source software packages on which we depend.
Individual groups or members have received support from AvH Foundation (Germany),
EPLANET, Marie Sk\l{}odowska-Curie Actions and ERC (European Union), 
Conseil G\'{e}n\'{e}ral de Haute-Savoie, Labex ENIGMASS and OCEVU, 
R\'{e}gion Auvergne (France), RFBR and Yandex LLC (Russia), GVA, XuntaGal and GENCAT (Spain), Herchel Smith Fund, The Royal Society, Royal Commission for the Exhibition of 1851 and the Leverhulme Trust (United Kingdom).



\addcontentsline{toc}{section}{References}
\setboolean{inbibliography}{true}
\bibliographystyle{LHCb}
\bibliography{main,LHCb-PAPER,LHCb-CONF,LHCb-DP,LHCb-TDR,custom-bib}

\newpage

\newpage
\centerline{\large\bf LHCb collaboration}
\begin{flushleft}
\small
R.~Aaij$^{40}$,
B.~Adeva$^{39}$,
M.~Adinolfi$^{48}$,
Z.~Ajaltouni$^{5}$,
S.~Akar$^{6}$,
J.~Albrecht$^{10}$,
F.~Alessio$^{40}$,
M.~Alexander$^{53}$,
S.~Ali$^{43}$,
G.~Alkhazov$^{31}$
% ...

{\footnotesize \it
% ...
$ ^{5}$Clermont Universit{\'e}, Universit{\'e} Blaise Pascal, CNRS/IN2P3, LPC, Clermont-Ferrand, France\\
$ ^{6}$CPPM, Aix-Marseille Universit{\'e}, CNRS/IN2P3, Marseille, France\\
% ...
$ ^{10}$Fakult{\"a}t Physik, Technische Universit{\"a}t Dortmund, Dortmund, Germany\\
% ...
$ ^{31}$Petersburg Nuclear Physics Institute (PNPI), Gatchina, Russia\\
% ...
$ ^{39}$Universidad de Santiago de Compostela, Santiago de Compostela, Spain\\
$ ^{40}$European Organization for Nuclear Research (CERN), Geneva, Switzerland\\
% ...
$ ^{43}$Nikhef National Institute for Subatomic Physics, Amsterdam, The Netherlands\\
% ...
$ ^{48}$H.H. Wills Physics Laboratory, University of Bristol, Bristol, United Kingdom\\
% ...
$ ^{53}$School of Physics and Astronomy, University of Glasgow, Glasgow, United Kingdom\\
% ...
}
\end{flushleft}
 
\newpage
%

%

\end{document}